\documentstyle[aps,twocolumn,epsfig]{revtex}
\begin{document}
\title{Ideal Gases in  Time-Dependent Traps}
\author{Georg M.\ Bruun and Charles W.\ Clark}
\address{Electron and Optical Physics Division, 
National Institute of Standards and Technology, US Department of Commerce,
Gaithersburg, Maryland 20899-8410}
\maketitle
\begin{abstract}
We investigate theoretically the properties of an ideal trapped gas in a 
time-dependent harmonic potential. Using a scaling formalism, we are able to
present simple analytical results for two important classes of experiments:
free expansion of the gas upon release of the trap; 
and the response of the gas to a harmonic modulation of the trapping 
potential is investigated. We present specific results relevant to
current experiments on trapped Fermions.
\end{abstract}
\section{Introduction}
Recently, impressive experimental results concerning the quantum degenerate regime 
of a dilute gas of trapped Fermionic atoms have been presented~\cite{DeMarcoScience}. 
By cooling $\sim 7\times 10^5$  $^{40}$K atoms to $0.5$ of the 
Fermi temperature, the onset of Fermi degeneracy was observed in the thermodynamic 
and in the scattering properties of the gas. Also, progress towards achieving 
fermi degeneracy for trapped $^6$Li \cite{Mewescon} has been presented.  Such a 
weakly-interacting, degenerate Fermi-Dirac gas provides 
a new platform for exploring fundamental quantum many-body physics.
Several theoretical results dealing with the equilibrium properties of such a gas 
have been presented~\cite{Butts,Schneider,BruunN}. For example, it has been shown 
that a two-component gas of spin-polarized $^6$Li undergoes a 
Bardeen-Cooper-Schrieffer (BCS) transition to superfluidity at experimentally 
obtainable densities and temperatures~\cite{StoofBCS,BruunBCS}. 

One class of experiments that is compatible with trapping
protocols, and which has gleaned valuable information on
the dynamics of dilute Bose-Einstein condensates (BECs), involves
monitoring the response of the gas to a change in trapping
potential.  For example, transient modulation of the trapping
potential induces free ringing of the gas, which in the case
of BECs led to a direct determination of low-lying regions
of the quasi-particle spectrum \cite{QP}.  Complete release
of the trapping potential enables one to view the free expansion
of the gas; this established early on the essential validity of the
time-dependent Gross-Pitaevski equation for describing the 
dynamics of latest generation of BECs \cite{HollandCooper},
which was later put to stringent quantitative tests~\cite{Expansion}. 

A trapped, single-component gas of ultracold Fermionic atoms can,
for experiments of current interest, be considered to be ideal 
(noninteracting) as a reasonable approximation, since
atomic collisions are strongly 
suppressed~\cite{DeMarco2,DeMarco}. 
In this paper, we examine theoretically the dynamics 
of such an ideal gas in a time-dependent trap. 
Using a scaling formalism similar to that which has
been successfully applied to trapped 
BECs~\cite{Castin,Kagan1,Kagan2,Dalfovo},
we derive analytical results for the
class of experiments mentioned above. 
The essential Fermionic character of the
systems with which we are concerned is established
by the initial distribution of particles in the
trap; the time evolution of this distribution, under
changes of the trapping potential that preserve
its harmonicity, are rigorously equivalent to that
of an ensemble of noninteracting particles, independent
of statistics.
We present a simple formula that describes
the free expansion of such an ideal gas, 
and our results suggest a new approach to the 
problem of quantitative thermometry in
the nanokelvin regime.  Due to their weak 
pair interactions, single-component Fermi-Dirac gases
are attractive candidates for nanokelvin thermometry;
with an improved theoretical understanding of finite-temperature
properties of BECs~\cite{BECT}, which are much more robust
candidates for experiment at present, we can envisage a direct
comparison of temperatures of ultracold Bose-Einstein and
Fermi-Dirac gases.  We also examine in detail the (non-linear) response 
of the gas to a harmonic oscillation of 
the trapping frequency. This reveals a domain of 
driving frequencies and amplitudes which 
generates a resonant response of the gas. 

The non-interacting limit considered here should provide a good approximation 
for degenerate, single-component cold Fermi-Dirac gases.
Since interactions in general play a smaller role for Fermi-Dirac vs. Bose-Einstein 
systems, the present approach may also provide a useful starting point for 
consideration of the dynamics of multiple-component Fermi gases.

\section{Formalism}
We start with a derivation of the equations describing the
scaling properties of an ideal gas trapped in a time-dependent 
harmonic potential. Consider a classical particle of mass $m$ trapped in 
a potential $V({\mathbf{r}},t)=m\sum_j\omega_j(t)^2 r_j^2/2$ with $j=x,y,z$ 
denoting the 3 spatial dimensions.  We set $\omega_j(t) = \omega_{0j}$
for $t \leq 0$, i.e. the trap potential is constant 
prior to $t =0$.  Newton's law is then expressed by 
$\dot{p}_j=-m\omega_j(t)^2 r_j$ and $m\dot{x}_j=k_j$. Invoking a
scaling 
transformation  $q_j=r_j/\gamma_j(t)$ and
$p_j=\gamma_j(t)k_j-\dot{\gamma_j}(t)mr_j$,
we obtain $\partial_{\tau_j} p_j=-m\omega_{0j}^2q_j^2$, with 
$\tau_j(t)=\int^t dt'\gamma_j(t')^{-2}$, if the scaling parameters
$\gamma_j(t)$ satisfy the equations 
\begin{equation}\label{gammaeq}
\ddot{\gamma}_j(t)=\frac{\omega_{0j}^2}{\gamma_j(t)^3}-\omega_j(t)^2\gamma_j(t).
\end{equation}
In quantum mechanics, the Heisenberg equation 
for a non-interacting gas in a time-dependent harmonic potential
takes the form:
\begin{equation}
i\hbar\partial_t\hat{\psi}({\mathbf{r}},t)=[-\frac{\hbar^2}{2m}\nabla^2
+V({\mathbf{r}},t)]\hat{\psi}({\mathbf{r}},t).
\end{equation}
where $\hat{\psi}({\mathbf{r}})$ is the field operator for an atom 
in a single hyperfine state at position ${\mathbf{r}}$, which obeys the 
usual Fermi-Dirac anticommutation relations. The quantum mechanical 
analogue of the rescaling outlined for the classical case 
above corresponds to writing~\cite{Castin}
\begin{equation}\label{transform}
\hat{\psi}({\mathbf{r}},t)=
\frac{\hat{\Phi}[{\mathbf{q}}(t)]}{\sqrt{\gamma_x\gamma_y\gamma_z}}
 e^{im\sum_jr_j^2\dot{\gamma}_j/2\hbar\gamma_j}
\end{equation}
with  $q_j(t)=r_j/\gamma_j(t)$, as defined above. If each 
 $\gamma_j$ satisfies Eq.(\ref{gammaeq}), by writing  
 $\hat{\Phi}[{\mathbf{q}}(t)]=\Pi_j\hat{\phi}[q_j(t)]$, we obtain 
\begin{equation}
\label{scaledSE}
i\hbar\partial_{\tau_j}\hat{\phi}[q_j(t)]=[-\frac{\hbar^2}{2m}\partial^2_{q_j}
+\frac{1}{2}m\omega_{0j}^2q_j^2]\hat{\phi}[q_j(t)].
\end{equation}
Thus, the time-dependent problem has been reduced 
to the trivial case of evolution in a 
time-independent harmonic trap in the rescaled variables $(\tau_j,q_j)$.
Determination of these variables requires only 
the solution of the three ordinary differential equations
expressed by Eq.(\ref{gammaeq}); the three spatial dimensions
can be treated independently.  An appealing qualitative
description of the solution of these equations, cast in the
context of one-dimensional scattering theory, has been given
by Kagan et al.\cite{Kagan1}. From Eq.(\ref{transform}), it 
follows that the density $\rho({\mathbf{r}},t)$ of the 
gas for a given time $t$
is given by 
\begin{equation}\label{density}
\rho({\mathbf{r}},t)\equiv\langle\hat{\psi}^\dagger({\mathbf{r}},t)
\hat{\psi}({\mathbf{r}},t)\rangle=\frac{1}{\gamma_x(t)\gamma_y(t)\gamma_z(t)}
\rho_0[{\mathbf{q}(t)}]
\end{equation}
where $\rho_0({\mathbf{r}})$ is the particle density for $t=0$. 
Thus, we can calculate $\rho({\mathbf{r}},t)$
for any time $t>0$ for modulated frequencies $\omega_j(t)$, 
by solving Eq.(\ref{gammaeq}), subject to the boundary conditions
$\gamma_j(0)=1$ and $\dot{\gamma}_j=0$.  Equation (\ref{gammaeq}) also describes the 
two-dimensional, non-ideal BEC subject to isotropic variations
of the trap potential, with Eq.(\ref{density}) also
being applicable if $\rho_0$ is obtained by a solution of the
Gross-Pitaevski equation~\cite{Kagan1}.

We now analyze the solution of Eq.(\ref{gammaeq})
for two cases of experimental relevance: free expansion; 
and harmonic modulation
of the trapping potential.

\section{Free Expansion} 
To model a free expansion experiment, we take 
$\omega_j(t)=0$ for $t>0$. 
Solving Eq.(\ref{gammaeq}) with 
the requirement that the gas is in equilibrium for $t\le 0$, we obtain
$\gamma_j(t)=1$ for $t\le 0$ and 
\begin{equation}\label{gammafree}
\gamma_j(t)=\sqrt{1+\omega_{0j}^2t^2}
\end{equation}
for $t > 0$. The time-dependent width of the cloud after the trap has 
been dropped is given by 
$\sqrt{\langle \hat{r}^2_j\rangle(t)}=\gamma_j(t)\sqrt{\langle
\hat{r}^2_j\rangle(0)}$.
To describe the aspect ratio $\alpha(t)$ of a cylindrically-symmetric
cloud, we find 
\begin{equation}\label{alpha}
\alpha(t)\equiv\sqrt{\frac{\langle x^2\rangle(t)}{\langle z^2\rangle(t)}}=
\alpha(0)\sqrt{\frac{1+\omega_{0x}^2t^2}{1+\omega_{0z}^2z^2}}
\end{equation}
We now apply these results to an ideal gas in 
two limiting cases. 

\begin{figure}
\centering
\epsfig{file=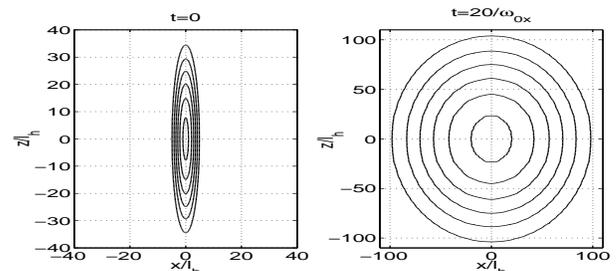,height=0.2\textwidth,width=0.45\textwidth,angle=0}
\caption{Contour images of the integrated density $\rho(x,z,t)$ for a freely
expanding gas. 
There are 9310 atoms in the cloud. We have defined 
$l_h\equiv(\hbar/m\omega_{0x})^{1/2}$.}
\label{Expandfig}
\end{figure}

\begin{figure}
\centering
\epsfig{file=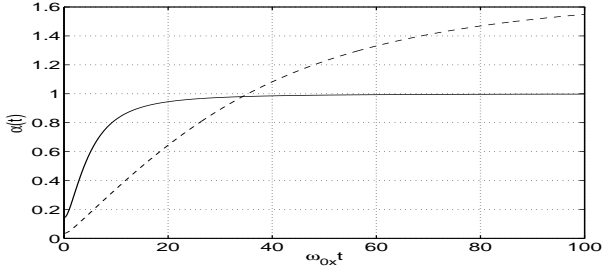,height=0.2\textwidth,width=0.45\textwidth,angle=0}
\caption{The aspect ratio for a freely expanding gas. The solid line is for 
$\lambda_z=19.5/137$, $\lambda_y=1$ and $\mu_F=20\hbar\omega_{0x}$. 
The dashed line is for $\lambda_z=\lambda_y=1/50$ and 
$\mu_F=\hbar\omega_{0x}$.}
\label{alphafig}
\end{figure}

We first treat the case of $T=0$, with the chemical potential
$\mu_F(T=0)/\hbar\omega_{0j}\gg 1$ for $j=x,y,z$. This corresponds
to the semiclassical limit, appropriate to current experiments
with more than a few hundred trapped atoms.  In this limit, the 
initial density profile is well described by the Thomas-Fermi (TF) 
approximation~\cite{Butts}. This gives the integrated density  
$\rho(x,z,t)\equiv\int dy \rho({\mathbf{r}},t)$,
\begin{equation}\label{densT0}
\rho(x,z,t)=\frac{m\mu_F}{4\pi\hbar^3\omega_{0y}\gamma_x\gamma_z}\left[1-
\frac{x^2/\gamma_x^2+\lambda_z^2z^2/\gamma_z^2}{R_F^2}\right]^2
\end{equation}
with $\lambda_z\equiv\omega_{0z}/\omega_{0x}$ and
$R_F=\sqrt{2\mu_F/m\omega_{0x}^2}$. 
We see that the cloud becomes isotropic for $\omega_{0j}t\gg 1$.  
This is the result one would get for a classical gas at finite
temperature, in which the momentum distribution is isotropic,
and is consistent with the initial isotropic momentum distribution
implicit in the TF approximation (in BECs in anisotropic traps, 
on the other hand, the long-time expansion is anisotropic due to 
macroscopic population of an anisotropic initial state, a key
result in their first observation~\cite{JILA1}).
The initial aspect ratio of $\alpha(0)=\lambda_z$ and Eq.(\ref{alpha})
evolves to $\alpha(t)\rightarrow 1$ as $t\rightarrow \infty$. 
In Fig.(\ref{Expandfig}), we plot the integrated density $\rho(x,z,t)$ for
$T=0$, $\mu_F(T=0)=20\hbar\omega_{0x}$, and $t=0$ (a) 
and $t=20/\omega_{0x}$ (b). We have taken  
$\lambda_z=19.5/137$ and $\lambda_y\equiv\omega_{0y}/\omega_{0x}=1$
corresponding to 
current experiments on trapped  $^{40}$K atoms~\cite{DeMarcoScience}.
 The two snapshots of the integrated density clearly 
show that the gas becomes isotropic for large expansion times. This is
further illustrated in Fig.(\ref{alphafig}), where we plot $\alpha(t)$ as given in 
Eq.(\ref{alpha}) for the same set of parameters. 
Note that  $\alpha(0)=\lambda_z$ for any initial density distribution 
of the form $\rho_0({\mathbf{r}})=f(\sum_j\omega_{0j}^2r_j^2)$. Such initial
distributions will become isotropic in an free expansion experiment. For
instance, for  $T>T_F\equiv\mu_F(T=0)/k_B$, the density is well described by a
classical gaussian profile, i.e.\ 
 $\rho_0({\mathbf{r}})\propto\exp[-\beta(\mu_F-m\sum_j\omega_{0j}^2r_j^2/2)]$
and Fig.(\ref{alphafig}) thus describes a trapped gas of fermions for any
$T$ within the TF approximation. This means that one cannot detect the 
onset of Fermi degeneracy by looking at the aspect ratio of the expanding gas.

As an example where $\rho_0({\mathbf{r}})\neq f(\sum_j\omega_{0j}^2r_j^2)$,
we now consider a case with $\mu_F/\hbar\omega_{0x}<3/2$ such that only one
level is occupied in the $x$-direction and $\mu_F/\hbar\omega_{0j}\gg 1$ for
$j=y,z$. The gas is initially strongly confined in the $x$-direction.
The integrated density profile is then
\begin{equation}
\rho(x,y,t)\propto e^{x^2/l_h^2\gamma_x^2}
\left(1-\frac{m\omega_{0y}^2y^2/\gamma_y^2}{2\mu_F-\hbar\omega_{0x}}\right)^
{3/2}
\end{equation}
for $t\ge0$, yielding
$\alpha(0)=\sqrt{3\hbar\omega_{0x}}\lambda_z/\sqrt{2\mu_F-\hbar\omega_{0x}}$.
In Fig.(\ref{alphafig}), we plot the aspect ratio for a free expansion for 
$\mu_F=\hbar\omega_{0x}$ and $\lambda_y=\lambda_z=1/50$ using Eq.(\ref{alpha}). 
We have  $\alpha(0)=\sqrt{3}/50$ and $\alpha(t)\rightarrow \sqrt{3}$ for 
$t\rightarrow \infty$. The gas, which initially 
is strongly confined in the $x$-direction will for $\omega_{0j}t\gg 1$
become most confined in the $y$-, and $z$-directions. This is, of course, a direct
reflection of the uncertainty relation giving higher average momentum in the
$x$-direction. However, observation of such a quantum effect requires a highly 
anisotropic trap. For the case of $10^4$ trapped atoms we would require  
$\lambda_y=\lambda_z\simeq 1/250$. Thus, for Fermi-Dirac particles 
the anisotropy of the expanded cloud due to the Heisenberg uncertainty principle 
is considerably less than that encountered in the Bose-Einstein case. 

From a measurement of the density at any time $t$ under free expansion 
it is straightforward to determine the initial density distribution 
$\rho_0({\mathbf{r}})$ using Eq.(\ref{density})-(\ref{gammafree}). This
suggests that a trapped, single-component Fermi-Dirac gas could 
serve as a low $T$ thermometer. Assuming the gas is in 
thermodynamic equilibrium for $t\le0$, one could infer the temperature of the gas 
from a knowledge of its density distribution,
the calculation of which is a trivial problem of summing over 
fractionally occupied trap levels. Hence, a determination 
of $T$ is simply a matter of fitting a calculated density to the 
measured $\rho({\mathbf{r}})$. Density profiles 
for an ideal Fermi-Dirac gas for various temperatures have already been given
elsewhere~\cite{Butts}.  However, if only one hyperfine level is 
present at all times, the atomic collisions required for thermalization are suppressed 
at low temperatures, and evaporative cooling of the gas 
will be inhibited. This can be remedied by initially trapping the gas in two 
hyperfine states, to allow $s$-wave collisions
between atoms in different states. One can then deplete the trap of one hyperfine
state by applying a microwave field that depletes 
one of the states. If this depletion occurs on a time scale 
that is long compared to collision times in the sample, the remaining 
atoms will be in thermodynamic equilibrium, and their expansion 
can be described by the results in this section. 
In fact, a procedure of this type is being used
in present experiments on trapped $^{40}$K atoms~\cite{DeMarcoScience}. 

Alternatively, one could drop the trap with both hyperfine levels present. The initial 
density profile and the free expansion of the gas will then be perturbed away from the 
ideal gas result by the interactions. When the effect of interactions is small, 
the initial equilibrium density and the subsequent expansion can be calculated 
relatively easily. A calculation of the effect of the 
interactions on the equilibrium distribution has been presented 
elsewhere,~\cite{BruunN} and the effect of the interactions on the 
free expansion can be calculated within the mean-field approximation, starting from 
the simple uncoupled solutions given by Eq.(\ref{gammafree}) for each spatial 
direction~\cite{Bruunexp}.

\section{Driven oscillations of the gas}
As noted above, modulation of the trapping frequency 
has proven to be a useful technique for probing 
the low-lying collective modes of trapped BECs. We therefore examine 
theoretically the non-linear response of a trapped 
gas to such a modulation. The non-interacting case 
treated here is in some sense opposite to
the hydrodynamic limit we have treated elsewhere~\cite{Bruunhydro}. 
To model a typical experiment, we assume that the trapping frequencies take the form 
 $\omega_j(t)^2=\omega_{0j}^2[1-2\eta\cos(\omega_Dt)]$
for $t>0$ with $\omega_D$ being the driving frequency and $\eta$ the driving
amplitude. Instead of solving Eq.(\ref{gammaeq}) with this form for $\omega_j(t)$, it
turns out to be  easier to go back to the original classical equation of motion 
 $\ddot{r}_j=-\omega_j(t)^2r_j$ by using 
 $r_j(t)=q_j(t)\gamma_j(t)$ and $q_j(t)=\exp[\pm i\omega_{0j}\tau(t)]$.
Thus, by writing $\xi_j(t)=\gamma_j(t)\exp[i\omega_{0j}\tau(t)]$, and 
 $\chi=\omega_Dt/2$, we obtain:
\begin{equation}\label{Mathieu}
\partial_\chi^2\xi_j+[a-2q\cos(2\chi)]\xi_j=0
\end{equation}
with $a=4/\tilde{\omega}^2$, $q=4\eta/\tilde{\omega}^2$ and 
$\tilde{\omega}=\omega_D/\omega_{0j}$. Equation (\ref{Mathieu}) is 
a variant of Mathieu's equation~\cite{AS}, whose properties
have been extensively studied.  Its behavior is easy to explain in one case
of experimental interest: that when the trap frequency is modulated
during a finite interval $0<t\le t_D$, and then returned to its original value.
The subsequent motion of the cloud is described by a solution
of the time-independent problem  $\omega_j(t)=\omega_{0j}$ 
with arbitrary initial conditions:
\begin{equation}\label{gammagen}
\gamma_j(t)=\sqrt{\sqrt{E^2-1}\sin(2\omega_{0j}t+c)+E},
\end{equation}
where $E=(\omega_{0j}^{-2}\dot{\gamma}_j^2+\gamma_j^2+\gamma_j^{-2})/2\ge 1$
is a conserved quantity for $t>t_D$; $c=\arcsin[(\gamma_{0j}^2-E)/\sqrt{E^2-1}]$; 
and $\gamma_{0j}=\gamma_j(t_D)$ is the value of $\gamma_j(t)$ immediately after the 
modulation ceases. Equation (\ref{gammagen}) has a discrete
frequency spectrum (frequencies of $2n\omega_{0j}$ with $n=0,1,2\ldots$),
as expected for a non-interacting gas 
in a harmonic trap. The linear response limit is recovered by letting 
 $E\rightarrow 1_+$ inEq.(\ref{gammagen}), to obtain 
 $\gamma_j(t)=1+\delta\sin(2\omega_{0j}t)$.

In essence, the problem of predicting the response of the 
gas to a harmonic driving with frequency $\omega_D$ and amplitude $\eta$ 
is reduced to an analysis of the well-known solutions to Mathieu's equation. 
In the parameter space $(a,q)$, there are regions where 
the solutions of Eq.(\ref{Mathieu}) are unstable, i.e.\ their amplitude increases
exponentially with time. Also, there are stable regions where the solutions remain 
bounded. The solutions on the boundaries between these regions are the Mathieu 
functions~\cite{AS}. Using $|\gamma_j(t)|=|\xi_j(t)|$, 
$a=4/\tilde{\omega}^2$, and $q=4\eta/\tilde{\omega}^2$, this means that for certain 
regions in the $(\omega_D,\eta)$-space, the response of the gas 
diverges as the driving time $t_D$ increases, i.e.
there is a resonant response, whereas in other regions the 
response of the gas remains finite for 
any value of $t_D$. Specifically, for $a=n^2$ with  $n=0,1,2\ldots$, the solutions to
Eq.(\ref{Mathieu}) diverge in time for an arbitrary small $q$~\cite{AS}. 
Thus, for
\begin{equation} 
 \omega_D=2\omega_{0j}/n,\ \   n=1,2,3\ldots
\end{equation}
 the response of the gas is resonant for an arbitrarely small driving
amplitude and  the amplitude of its oscillations will diverge with the driving time. 
 In terms of Eq.(\ref{gammagen}), if the gas remains trapped in the
time-independent potential $\omega_j(t)=\omega_{0j}$ 
after the driving ($t>t_D$), we have $E\gg 1$ and the oscillations of the gas will 
be large and contain many harmonics. The resonance for $\omega_D=2\omega_{0j}$ is, 
of course, the usual excitation frequency for an even-parity perturbation for a
non-interacting gas. However, the resonances for 
 $\omega_D=\omega_{0j}, 2\omega_{0j}/3, \omega_{0j}/2$ etc.\ do not correspond 
to new modes. They simply reflect the fact that for these driving 
frequencies, the trapping potential is doing resonant work on the gas. Note that the 
present scaling formalism does not predict the higher-frequency modes at 
 $n2\omega_{0i}$ with $n\ge 2$, as they do not correspond to simple 
dilations/contractions of the  cloud. 

\begin{figure}
\centering
\epsfig{file=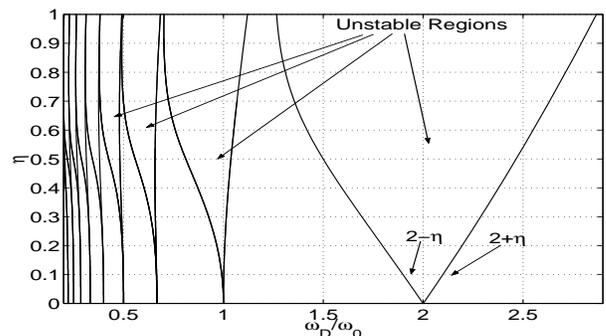,height=0.25\textwidth,width=0.45\textwidth,angle=0}
\caption{The regions of stability/unstability for the response of a non-interacting 
gas to a modulation of the trapping potential with frequency $\omega_D$ and 
amplitude $\eta$.}
\label{Mathieufig}
\end{figure}

We will now examine the width of the unstable 
(resonant) regions around $\omega_D=2\omega_{0j}/n$ for $\eta\rightarrow 0$.
For the $\omega_D=2\omega_{0j}$ resonance, the unstable region of the
Mathieu equation  is bounded by $1-q<a<1+q$ for $q\rightarrow 0$~\cite{AS}. This 
gives, that for driving frequencies and amplitudes such that 
\begin{equation}
2-\eta<\tilde{\omega}<2+\eta,\ \ \eta\ll 1, 
\end{equation} 
the response of the gas is divergent. Hence, the resonance region of the gas for 
finite  but small driving amplitude $\eta$ has a reasonable width and should be
relatively easy to access experimentally. Likewise, for the $\omega_D=\omega_{0j}$
resonance, the  unstable region is bounded by  
\begin{equation}
1-5\eta^2/6<\tilde{\omega}<1+\eta^2/6,\ \ \eta\ll 1
\end{equation}
and it should be experimentally accessible. For the lower frequency
resonances  ($\omega_D=2\omega_{0j}/n,\ n\ge 3$), it turns out that the resonance
regions  are very narrow for $\eta\rightarrow 0$ as the boundary lines between 
the stable and unstable solutions of Eq.(\ref{Mathieu}) only differ by terms of 
order $q^3$ or higher~\cite{AS}. Thus, these resonances are difficult to find 
experimentally for $\eta\rightarrow 0$. Instead, one could increase the driving 
amplitude $\eta$ for a given frequency $\omega_D$; for large enough $\eta$, an  
unstable region is reached and the response of the gas diverges. The above results 
are illustrated in Fig.(\ref{Mathieufig}), where we plot the response of the gas 
as a function of driving frequency $\omega_D$ and amplitude $\eta$.
 The lines in the plot separate regions where the response of the gas
remains finite from regions where the oscillations of the gas diverge with driving 
time. For example, if one modulates the external potential with an off-resonance
frequency of  $\tilde{\omega}=5/3$, the response of the gas remains finite for small
$\eta$. However, if one gradually increases the driving amplitude $\eta$ for this
frequency, then one should observe a divergent response of the gas for $\eta$ larger
than  $\sim 0.33$. The plot is  generated using the well-known properties of the 
Mathieu equation  and the mapping 
$(a,q)\rightarrow(4/\tilde{\omega}^2,4\eta/\tilde{\omega}^2)$. Note the unstable 
(resonance) regions for $\eta\rightarrow 0$ for $\omega_D=2\omega_{0j}/n$. The 
regions have zero width for $\eta\rightarrow 0$ for $n\ge3$ as predicted above. 
For increasing driving amplitude $\eta$, the unstable regions grow as expected. 
For  $\omega_D\rightarrow 0$, the response of the gas for $\eta<\sim0.5$ is
finite whereas it diverges for $\eta>\sim0.5$. Physically, this corresponds to the
fact that for $\omega_D/\omega_{0i}\ll1$ and $\eta<0.5$, the gas returns
adiabatically to its initial state after one period of trap modulation and there is no 
net work done. Contrary, for $\eta>0.5$ the trapping potential becomes inverted for 
certain times resulting in a divergent response of the gas. Mathematically,
the transition region for $\eta\simeq 0.5$ comes from the fact, that for $a\sim
2q$, the even and odd periodic solutions (Mathieu functions) of Eq.(\ref{Mathieu})
start to differ in "energy" (the Mathieu characteristic value $a$~\cite{AS}), as the
tunneling between successive minima of the potential potential  $\cos(2\chi)$ becomes
significant. 

Again, since a spin-polarized gas of cold fermionic atoms is effectively
ideal, the results presented here should apply to the response of such a gas to 
a harmonic modulation of the trapping potential.

\section{Conclusion}
In conclusion, using a scaling formalism we have been able to derive simple
analytical results for the dynamics of ideal gases trapped in time-dependent harmonic 
traps. We have concentrated on two important classes of experiments: 
For free expansion, we show how the initial density profile of 
the gas can easily be determined from a measurement of the density profile at any 
time $t$ after the trap has been dropped. We proposed a low $T$ thermometer based on 
these results. Also, we showed how the problem of determining non-linear response of 
the gas to a harmonic modulation of the trapping frequency can be  mapped to 
an analysis of the well-known properties of the solutions to Mathieu's equation. 
We identified regions in $(\omega_D,\eta)$-space where the response of the gas was 
divergent with the modulation time. Especially, we were able to predict unstable 
regions for $\eta\rightarrow 0$ reflecting the fact, that the trap is doing resonant 
work on the gas. Since ultracold spin-polarized fermions are non-interacting to a 
very good approximation, our results should  be directly relevant for 
current experiments in this very active field of research.
\section{Acknowledgements}
We thank Y.\ Castin and D.\ Jin for useful discussions.

\end{document}